\documentclass{article}

\usepackage[preprint,nonatbib]{neurips_2022}

\usepackage[utf8]{inputenc} % allow utf-8 input
\usepackage[T1]{fontenc}    % use 8-bit T1 fonts
\usepackage{hyperref}       % hyperlinks
\usepackage{url}            % simple URL typesetting
\usepackage{booktabs}       % professional-quality tables
\usepackage{amsfonts}       % blackboard math symbols
\usepackage{nicefrac}       % compact symbols for 1/2, etc.
\usepackage{microtype}      % microtypography

\usepackage{times}
\usepackage{latexsym}

\usepackage[T1]{fontenc}
\usepackage[utf8]{inputenc}

\usepackage{microtype}

\usepackage{inconsolata}
\usepackage{graphicx}
\usepackage{multirow}
\usepackage{booktabs}

\title{InPars-v2: Large Language Models as Efficient Dataset Generators for Information Retrieval}

\author{Vitor Jeronymo\\
NeuralMind, Brazil\\
FEEC-UNICAMP, Brazil \\\And
Luiz Bonifacio \\
NeuralMind, Brazil\\
FEEC-UNICAMP, Brazil \\ \And
Hugo Abonizio\\
NeuralMind, Brazil\\
FEEC-UNICAMP, Brazil \\\AND
Marzieh Fadaee \\
Zeta Alpha, Netherlands \\\And
Roberto Lotufo\\
NeuralMind, Brazil\\
FEEC-UNICAMP, Brazil \\\And 
Jakub Zavrel \\
Zeta Alpha, Netherlands \\\And
Rodrigo Nogueira \\
NeuralMind, Brazil \\
FEEC-UNICAMP, Brazil \\
Zeta Alpha, Netherlands
}

\begin{document}
\maketitle


%% the authors and 
% their affiliations.

%\authorrunning{Jeronymo et al.}
%\titlerunning{InPars-v2}

%\maketitle              % typeset the header of the contribution

\begin{abstract}
  Recently, InPars introduced a method to efficiently use large language models (LLMs) in information retrieval tasks: via few-shot examples, an LLM is induced to generate relevant queries for documents. These synthetic query-document pairs can then be used to train a retriever. However, InPars and, more recently, Promptagator, rely on proprietary LLMs such as GPT-3 and FLAN to generate such datasets. In this work we introduce InPars-v2, a dataset generator that uses open-source LLMs and existing powerful rerankers to select synthetic query-document pairs for training. A simple BM25 retrieval pipeline followed by a monoT5 reranker finetuned on InPars-v2 data achieves new state-of-the-art results on the BEIR benchmark. To allow researchers to further improve our method, we open source the code, synthetic data, and finetuned models: \url{https://github.com/zetaalphavector/inPars/tree/master/legacy/inpars-v2}
\end{abstract}

\section{Introduction and Background}

Data augmentation has been a reliable tool to improve the effectiveness of AI models in the face of the scarcity of high-quality in-domain training data, which is a common problem in practical applications.
Previous work by Bonifacio et al.~\cite{bonifacio2022inpars} and Dai et al.~\cite{dai2022promptagator} successfully leveraged the few-shot capabilities of LLMs to generate reliable synthetic training data for information retrieval models. These training data helped their models achieve state-of-the-art (SOTA) results on the BEIR benchmark~\cite{beir}.

Bonifacio et al.~\cite{bonifacio2022inpars} propose InPars where they generate queries from documents in the corpus using LLMs. 
Similarly to Bonifacio et al.~\cite{bonifacio2022inpars}, the recently published Promptagator~\cite{dai2022promptagator} model also feeds prompts to LLMs in order to generate alternative queries for a given document in an unsupervised manner. It differs primarily from InPars in that it uses dataset-specific prompts, a larger LLM to generate queries, and a fully trainable retrieval pipeline with smaller models.

This work extends the method of Bonifacio et al.~\cite{bonifacio2022inpars} by using a reranker as a filtering mechanism to select the best synthetically generated examples and further improving retrieval effectiveness on BEIR. We also use an open-source query generator as opposed to the proprietary one used by Bonifacio et al. and provide the source code and data to reproduce our results on TPUs.
We refer to Bonifacio et al.~\cite{bonifacio2022inpars} model as Inpars-v1 and the model presented in this paper as Inpars-v2.

%In comparison to InPars-v1, InPars-v2 uses an open-source query generator as opposed to the proprietary one used in v1 and an improved method to select the best synthetic data generated by the LLM to fine-tune the reranker, thus improving the overall retrieval effectiveness.

% Most recently, LLMs such as GPT-3~\cite{brown2020language} have been used for Natural Language now in a variety of tasks, such as Text Classification ~\cite{sahu2022data,bayer2022multi,nouri-2022-data}, Information Retrieval ~\cite{bonifacio2022inpars,dai2022promptagator}, Question Answering ~\cite{Sachan2022QuestionsAA}, and many more.

% This work contemplates, but does not focus entirely on generating new data as in InPars-v1: It rather uses an improved method to select the best synthetic data generated by the LLM to fine-tune the reranker and thus improving the overall retrieval effectiveness.

\section{Methodology}

In this section, we explain the experiments we performed and how they differ from InPars-v1~\cite{bonifacio2022inpars}.

% To standardize, streamline, and automate the evaluation and training pipelines, we used Pyserini's~\cite{lin2021pyserini} flat indexes\footnote{As opposed to the multifield index.} to generate BM25 runs for each dataset. To compare with previous work with this new pipeline, we reproduced the experiments as in Bonifacio et al.~\cite{bonifacio2022inpars} by reranking these runs with monoT5~\cite{nogueira2020document} with 3 billion parameters, finetuned on MS MARCO for one epoch\footnote{https://huggingface.co/castorini/monot5-3b-msmarco-10k}.

To generate synthetic queries, we use the open-source GPT-J~\cite{gpt-j} with 6B parameters to replace OpenAI's curie model used in InPars-v1.
For each dataset in the BEIR benchmark, we sample 100k documents from its corpus and generate one synthetic query per document using GPT-J prompted with 3 examples from MS MARCO. We use greedy decoding and the ``gbq'' prompt template from InPars-v1.
Some corpora in BEIR such as ArguAna~\cite{wachsmuth-etal-2018-retrieval} have less than 100k documents. In these cases, we generate as many synthetic queries as there are documents in the corpus.
It takes on average 30 hours on an A100 GPU to generate 100k queries.

Once the synthetic queries are generated, we apply a filtering step to select query-document pairs that are more likely to be relevant to each other. In InPars-v1, this filtering step consisted of selecting the top 10k query-document pairs with the highest log probabilities of generating a query given the 3-shot examples and the document as input.
In InPars-v2, we use monoT5-3B~\cite{nogueira2020document} already finetuned on MS MARCO for one epoch\footnote{https://huggingface.co/castorini/monot5-3b-msmarco-10k} to estimate a relevancy score for each of the 100k query-document pairs. Then, we keep only the top 10k pairs with the highest scores as our positive query-document pairs for training. It takes approximately 1.5 hours to score 100k query-document pairs on a TPU v3-8. It should take twice as much on a A100.

To obtain negatives (i.e., non-relevant) query-document pairs, we randomly sample one document from the top 1000 retrieved by BM25 when issued the synthetic query. Thus, our training set consists of 10k positive query-document pairs and 10k negative query-document pairs.

The rerankers are finetuned in the same manner as in InPars-v1: monoT5-3B is finetuned on MS MARCO for one epoch and then further finetuned  for one epoch on the synthetic data. We use the Adafactor optimizer~\cite{shazeer2018adafactor} with a constant learning rate of 1e-3. Each batch has 64 positive and 64 negative query-document pairs randomly sampled from the training dataset. We finetune one model on each synthetic dataset from BEIR, that is, we end up with 18 different rerankers, one per dataset, which are then evaluated on the corresponding test sets. Finetuning on each synthetic dataset takes less than 10 minutes on a TPU v3-8.

Evaluation is performed using the following pipeline: first we use Pyserini's~\cite{lin2021pyserini} flat indexes\footnote{As opposed to the multifield index.} to retrieve a thousand documents for each query using BM25 with default parameters (k1=0.9, b=0.4), for each dataset. Then we use the finetuned monoT5-3B models to rerank these documents.

\section{Results}

\begin{table*}[ht]
\centering
\begin{tabular}{l|c|ccc|cc}
\toprule
\multicolumn{1}{l|}{} &
  \multirow{2}{*}{\textbf{BM25}} &
  \multicolumn{3}{c|}{\textbf{monoT5-3B}} &
  \multicolumn{1}{l}{\multirow{2}{*}{\textbf{PrGator}}} &
  \multicolumn{1}{l}{\multirow{2}{*}{\textbf{RankT5}}} \\
\multicolumn{1}{l|}{} &
   &
  \textbf{MARCO} &
  \textbf{+InPars-v1} &
  \textbf{+InPars-v2} &
  \multicolumn{1}{l}{} &
  \multicolumn{1}{l}{} \\ \midrule
TREC-Covid    & 0.594 & 0.801 & 0.846 & 0.846 & 0.762 & 0.823 \\
Robust        & 0.407 & 0.615 & 0.610 & 0.632 & -     & -     \\
FiQA          & 0.236 & 0.509 & 0.492 & 0.509 & 0.494 & 0.493 \\
DBPedia       & 0.318 & 0.472 & 0.494 & 0.498 & 0.434 & 0.459 \\
SciDocs       & 0.149 & 0.197 & 0.206 & 0.208 & 0.201 & 0.191 \\
SciFact       & 0.678 & 0.774 & 0.774 & 0.774 & 0.731 & 0.760 \\
NFCorpus      & 0.321 & 0.383 & 0.385 & 0.385 & 0.370 & 0.399 \\
BioASQ        & 0.522 & 0.566 & 0.607 & 0.595 & -     & 0.579 \\
Natural Questions            & 0.305 & 0.625 & 0.625 & 0.638 & -     & 0.647 \\
HotpotQA      & 0.633 & 0.760 & 0.790 & 0.791 & 0.736 & 0.753 \\
TREC-News     & 0.395 & 0.477 & 0.458 & 0.490 & -     & -     \\
Quora         & 0.788 & 0.835 & 0.874 & 0.845 & -     & 0.819 \\
FEVER         & 0.651 & 0.848 & 0.852 & 0.872 & 0.866 & 0.848 \\
Climate-FEVER & 0.165 & 0.288 & 0.287 & 0.323 & 0.241 & 0.275 \\
Signal        & 0.328 & 0.302 & 0.319 & 0.308 & -     & 0.319 \\
ArguAna       & 0.397 & 0.379 & 0.371 & 0.369 & 0.630     & 0.406 \\
Touche        & 0.442 & 0.309 & 0.260 & 0.291 & 0.381 & 0.486 \\
CQADupstack   & 0.302 & 0.449 & 0.449 & 0.448 & -     & -     \\ \midrule
\textbf{Avg}           & 0.424 & 0.533 & 0.539 & 0.545 & -     & -     \\
\textbf{Avg PrGator}           & 0.417 & 0.520 & 0.523 & 0.533 & 0.531  & 0.536    \\ \bottomrule
\end{tabular}
\vspace{0.1cm}
\caption{nDCG@10 on BEIR. ``Avg PrGator'' is the average of datasets reported by Promptagator.}
\label{tab:main-results}
\end{table*}

Table~\ref{tab:main-results} presents results for BM25 (2nd column), monoT5-3B finetuned on MS MARCO (3rd column), monoT5-3b finetuned on MS MARCO and further finetuned on InPars-v1 (4th column), and monoT5-3B finetuned on MS MARCO and then finetuned on InPars-v2 data (5th column). Compared to InPars-v1, our approach is substantially better on TREC-News, Climate-FEVER, Robust and Touche. Additionally, we compare our method with Promptagator~\cite{dai2022promptagator} and RankT5~\cite{zhuang2022rankt5}. Taking into account the average of all BEIR datasets, these results represent a new state of the art on BEIR.

Promptagator and RankT5 strive on datasets that monoT5 and InPars-v2 cannot even surpass BM25, such as Touche and ArguAna. Note that these datasets focus on argument retrieval, which is slightly different from other datasets in the BEIR benchmark. As a result, they benefit from using custom prompts.\footnote{In preliminary experiments, we also observed an improvement of more than 10 nDCG@10 points on ArguAna by using a dataset-specific prompt to generate synthetic queries. More details and results on the full BEIR benchmark will appear in an upcoming paper.} Promptagator does this without using supervised data from MS MARCO and using smaller T5 models with 110M parameters for the retrieval and reranking steps.

Promptagator uses a proprietary model, FLAN~\cite{wei2021finetuned}, to generate synthetic queries. The RankT5 model is a modified version of the monoT5 reranker, but its checkpoint and code are not published. In this work, we make the code, models, and data open-source and publicly available. %On the BEIR benchmark on average, our method performs on par with state-of-the-art models.

% Note: add table here comparing curie vs gpt-j on 5 robust04 queries:
% gs://project-1462/maritaca/data/synthetic_robust04_curie_100k.jsonl

\section{Conclusion}

In this work, we presented InPars-v2, an improved version of InPars~\cite{bonifacio2022inpars} that uses a publicly available language model to generate queries and a better query-document pair selection process. Our results show that we achieve effectiveness on par with the state of the art on BEIR. The synthetic data and finetuned models were publicly released.

\section*{Acknowledgments}

This research was partially supported by Fundação de Amparo à Pesquisa do Estado de São Paulo (FAPESP) (project id 2022/01640-2). We also thank Centro Nacional de Processamento de Alto Desempenho (CENAPAD-SP) and Google Cloud for computing credits.  

\bibliographystyle{abbrv}
\bibliography{main}

%\pagebreak 

\onecolumn

%\pagebreak 
%\appendix

\end{document}